\documentclass[manuscript,trackchanges]{aastex62}
\usepackage{amsmath}
\usepackage{soul}

\hypersetup{linkcolor=red,citecolor=blue,filecolor=blue,urlcolor=blue}
\newcommand{\Rsun}{$R_{\odot}~$}
\newcommand{\Alfven}{Alfv\'{e}n~}
\shorttitle{Solar Coronal Magnetic Fields and Sensitivity of VELC onboard Aditya-L1}
\shortauthors{Sasikumar Raja et al.}
\begin{document}

\title{Solar Coronal Magnetic Fields and Sensitivity Requirements for Spectropolarimetry Channel of VELC Onboard Aditya-L1}

\correspondingauthor{K. Sasikumar Raja}
\email{sasikumar.raja@iiap.res.in}

\author[0000-0002-1192-1804]{K. Sasikumar Raja}
\affil{Indian Institute of Astrophysics, II Block, Koramangala, Bangalore-560 034, India}

\author{Suresh Venkata}
\affiliation{Indian Institute of Astrophysics, II Block, Koramangala, Bangalore-560 034, India}

\author{Jagdev Singh}
\affiliation{Indian Institute of Astrophysics, II Block, Koramangala, Bangalore-560 034, India}

\author{B. Raghavendra Prasad}

\affiliation{Indian Institute of Astrophysics, II Block, Koramangala, Bangalore-560 034, India}

\begin{abstract}

Understanding solar coronal magnetic fields is crucial to address the long-standing mysteries of the solar corona and solar wind. Although routine photospheric magnetic fields (MFs) are available for decades, coronal MFs are rarely reported. Visible Emission Line Coronagraph (VELC) on board Aditya-L1 mission (planned to launch in the near future) can directly measure the MFs in the inner solar corona. This can be achieved with the help of spectropolarimetric observations of the forbidden coronal emission line centered at 1074.7 nm over a field of view 1.05 R$_{\odot}$ - 1.5 R$_{\odot}$. In this article, we summarize various direct and indirect techniques used to estimate the MFs at different wavelength regimes. Further, we summarize the expected accuracies that are required to estimate MFs using VELC's spectropolarimetry channel.

\end{abstract}

%% Keywords should appear after the \end{abstract} command. 
%% See the online documentation for the full list of available subject
%% keywords and the rules for their use.
%% From the front matter, we move on to the body of the paper.
%% Sections are demarcated by \section and \subsection, respectively.
%% Observe the use of the LaTeX \label
%% command after the \subsection to give a symbolic KEY to the
%% subsection for cross-referencing in a \ref command.
%% You can use LaTeX's \ref and \label commands to keep track of
%% cross-references to sections, equations, tables, and figures.
%% That way, if you change the order of any elements, LaTeX will
%% automatically renumber them.
%%
%% We recommend that authors also use the natbib \citep
%% and \citet commands to identify citations.  The citations are
%% tied to the reference list via symbolic KEYs. The KEY corresponds
%% to the KEY in the \bibitem in the reference list below. 

\section{Introduction} \label{sec:intro}
%\S \textbf{Introduction \\ }
The dominant magnetic field (MF) in the solar corona plays a significant role in the formation, evolution, and dynamics of the structures in the solar corona. The MF is a crucial parameter in heating the corona to $\gtrsim 1$ MK. Transient events like solar flares (SFs), coronal mass ejections (CMEs), coronal jets, and other features like sunspots, filaments, prominences, polar plumes, and coronal holes, etc., are caused by the MF. After flare eruption, MFs channelize the particles to interplanetary medium (IPM). The accelerated or trapped electrons after the solar flares or MHD shocks driven by CMEs or SFs play a vital role in the origin of solar radio bursts (SRBs) at the meter, deca-hectometer, and kilometer wavelengths \citep{Sas2014, Dayal2019, Nda2021, Umi2021, Eoin2021}.

George Ellery Hale measured the magnetic field (MF) for the first time using Zeeman effect \citep{Hale1908}. At photosphere heights, the MF of the sunspots is estimated using the Zeeman effect. Nevertheless, direct MF measurements (using Zeeman effect and Hanle effects) are rare in low-density coronal plasma because of lack of suitable spectral lines, higher coronal temperature and weaker magnetic fields \citep{Vema2013, Chen2018}. Direct measurement of the magnetic field in the corona is difficult, and some of the estimated MFs have large uncertainties \citep{Lin2000, Lin2004, Harvey1969, Kuhn1995}. MFs estimated using Zeeman effect based on the `green', Fe XIV ($\lambda=530.3$ nm) observations is $13\pm20$ G \citep{Harvey1969}. Using spectro-polarimetric observations at near-infrared line Fe XIII ($\lambda = 1074.7$ nm), the estimated MF is 40 G \citep{Kuhn1995}. Linear polarization measurements of the Hanle effect in coronal emission line are successful in making maps of the direction of coronal MFs \citep{Mickey1973,Querfeld1984, Arnaud1987}. However, measurement using the Hanle effect is not sensitive to quantify the magnetic field strength \citep{Casini1999, Lin2000}. In recent years, it has been attempted to use forward calculations using FORWARD tool set \citep[for example see][]{Gib2016} to find a suitable magnetic field model by comparing the calculated and observed Stokes profiles to estimate the coronal magnetic fields \citep{Bak2013, Jib2016, Gib2017, Chen2018}.

Furthermore, there are some indirect methods to estimate the MFs in the inner corona. One of them is the radio polarization studies; as solar radio emissions at meter wavelengths are circularly polarized, it is possible to estimate MFs \citep[for e.g.,][]{Sas2013, Sas2014, Har2014}. The origin of the SRBs in the solar corona and IPM is due to different emission mechanisms like thermal bremsstrahlung, cyclotron emission, gyrosynchrotron, plasma emissions, etc \citep{Dulk1985}. Apart from these, band-splitting of type II bursts \citep{Kis2016, Kis2017}, quasi-periodicity of the Type III bursts \citep{Sas2013}, shock-standoff distance method using white light coronagraphs \citep{Gop2011, Kumari2017}, and EUV observations are used to estimate the MFs \citep{Gop2012, Kumari2019}. In the outer corona, it is possible to estimate the magnetic field using the Faraday rotation of linearly polarized radio signals emitted by natural sources, interplanetary space probes, and spacecraft radio beacons during their occultation by the solar corona \citep{Stelzried1970, Bird1981, Bird1982}.

Since routinely observed photospheric magnetograms are available, the extrapolation techniques can be used to estimate the MFs using different models. For instance, potential field source surface (PFSS) model \citep{Schatten1969, Sch2003, Sas2019} can be used to extrapolate the MF. PFSS model calculates the MF of the corona from 1 - 2.5 $R_{\odot}$ assuming that there are no currents in this region and thus satisfies $\nabla \times B = 0$, where B is the magnetic field.  Alternatively, `force-free' magnetic models which assume that MFs dominate all other forces (for e.g. plasma pressure and gravity) present in the chromosphere and corona \citep{Wie2012, Wie2014} can be used. Note that computation of latter case subdivides the problem into two cases: Linear Force Free Fields (LFF) which is mathematically simpler, and Non-Linear Force Free Fields (NLFFF) which is a general case but complex.

In this article, we summarize coronal MFs that are derived using various techniques (both direct and indirect) at different wavelengths in the field-of-view (FOV) of the VELC which are vital in planning the spectropolarimetric observations. 

\section{VELC}

\begin{figure*}[!ht]
	\centerline{\includegraphics[scale=0.8]{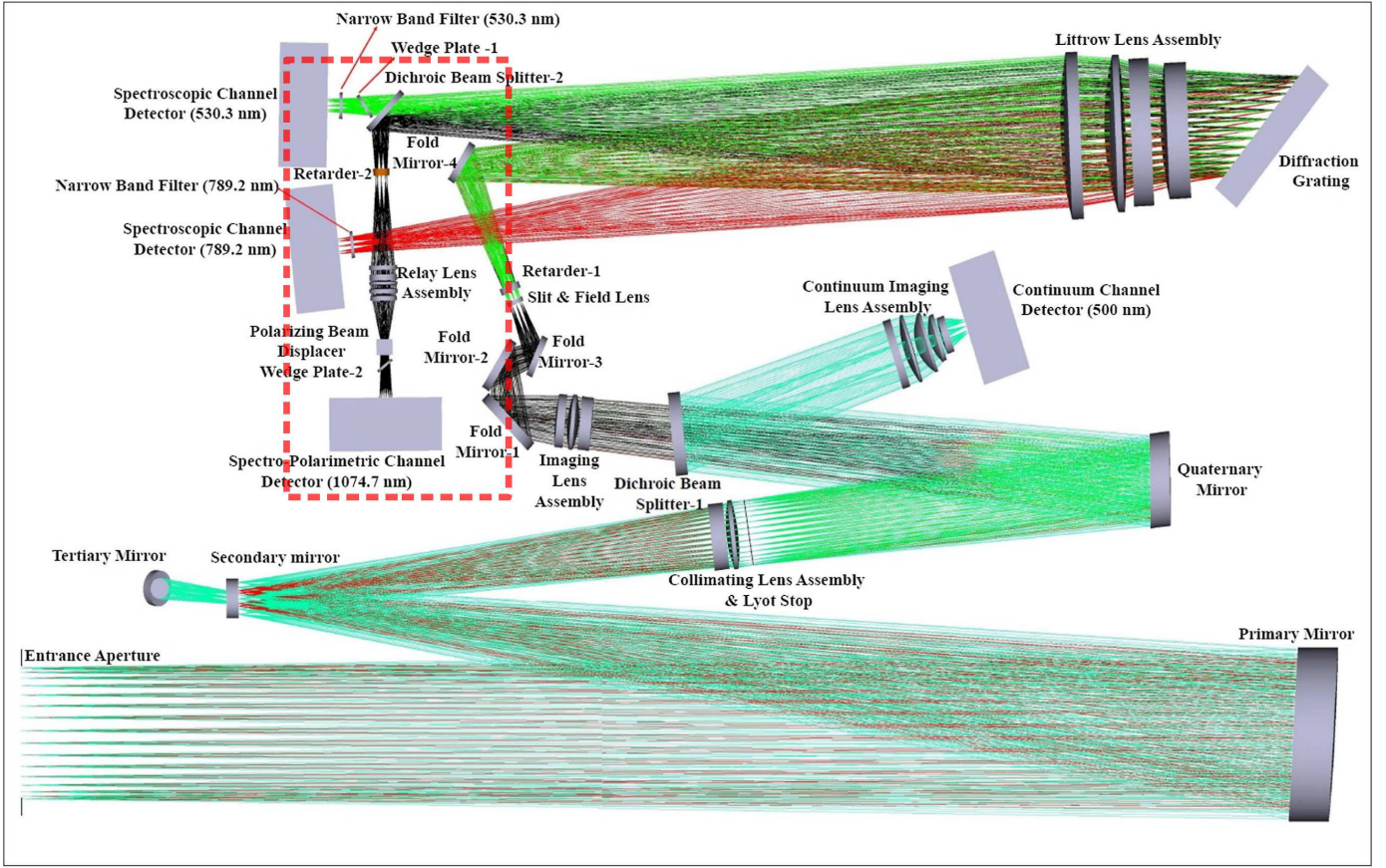}}
	\caption{Optical layout of Visible Emission Line Coronagraph on board Aditya-L1. The red color dashed box shows the spectropolarimetry channel.}
	\label{fig:velc}
\end{figure*}

Visible Emission Line Coronagraph (VELC) on board Aditya-L1 mission \citep{Seetha2017} is an internally occulting coronagraph. It is capable of observing solar corona in three different modes simultaneously: (i) continuum imaging of solar corona at 500 nm over a FOV of 1.05 - 3 $R_{\odot}$, (b) simultaneous multi-slit spectroscopic observations in coronal emission lines centered at 530.3 nm [Fe {\sc XIV}] and 789.2 nm [Fe {\sc XI}] and 1074.7 nm [Fe {\sc XIII}] over the FOV of 1.05 - 1.5 $R_{\odot}$, and (c) multi-slit dual-beam spectropolarimetry at Fe {\sc XIII} 1074.7 nm over the FOV of 1.05 - 1.5 $R_{\odot}$ \citep{Prasad2017,Raj2018,Singh2019}. For the sake of completeness optical layout of VELC is shown in Figure \ref{fig:velc}. 

To achieve the proposed scientific goals of the payload, total instrumental background over the FOV of the continuum channel is $\approx$ 5 ppm (parts per million) with respect to the disc brightness. VELC being an internally occulted reflective coronagraph, the major source of instrumental background is the scatter from the primary mirror due to surface microroughness and contamination from the other mirror surfaces. In order to understand the effects of different sources of scatter and to define the allowable limits for surface microroughness and contamination over the primary mirror surface, theoretical simulations and experiments are carried out by \citet{SV2017,SV2020,Narra2020}. %Complete details of these studies are discussed elsewhere \citep{SV,SV2020,narra}.

%\subsection{Spectropolarimetry channel}

\begin{figure*}[!ht]
	\centerline{\includegraphics[width=19cm]{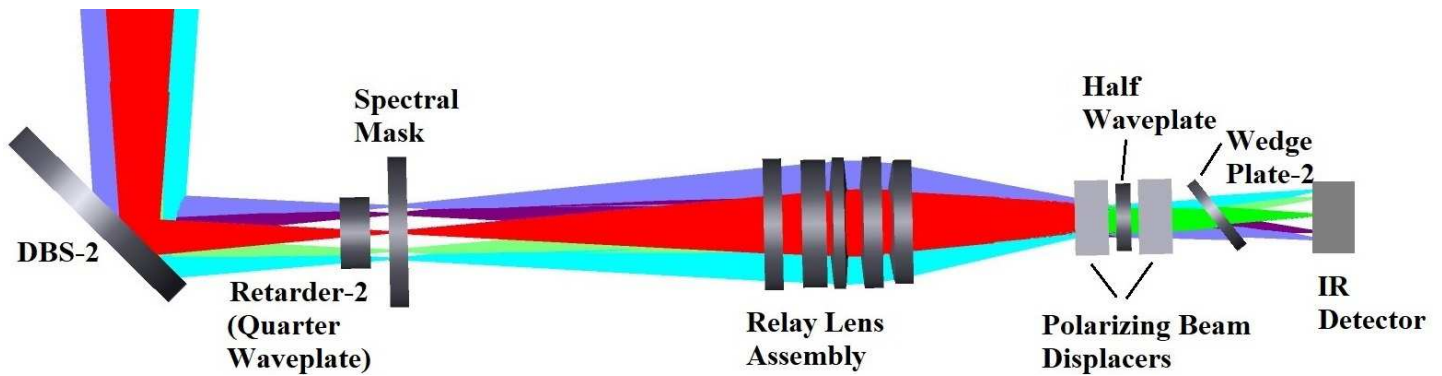}}
	\caption{Optical layout of spectropolarimetry channel of VELC.}
	\label{fig:sp}
\end{figure*}

Spectropolarimetry channel of VELC consists of Dichroic Beam Splitter-2 (DBS-2), which reflects the 1074.7 nm wavelength and transmits 530.3 nm wavelength. Reflected beam from DBS-2 will be transmitted through the retarder-2 (QWP; Quarter Wave Plate). Retarder-2 is mounted on a rotational mechanism with a rotation speed of $\approx 6$ RPM. Spectral mask placed at the intermediate focal plane after the retarder-2 helps in minimizing the spectral overlapping from the adjacent slits. Relay lens assembly re-images the intermediate focal plane on to the IR detector. In the converging beam path, a combination of Polarizing Beam Displacer (PBD), Half Wave Plate (HWP) and another Polarizing Beam Displacer (PBD) are placed. PBD -- HWP -- PBD combination acts as the analyzer package. Each PBD will provide half of the required separation between the O \& E rays ($\approx$ 1.875 mm). The HWP mounted after the first PBD flips the polarization states of the output beams from PBD-1 (i.e., O and E rays are flipped as E \& O rays). This ensures no path difference between the O \& E rays at the final focal plane. Wedge plate-2 is used to minimize the aberration produced due to PBD -- HWP -- PBD combination. Optical layout of the spectropolarimetry channel of VELC is shown in Figure \ref{fig:sp}.

\begin{figure*}[!ht]
	\centerline{\includegraphics[width=10cm]{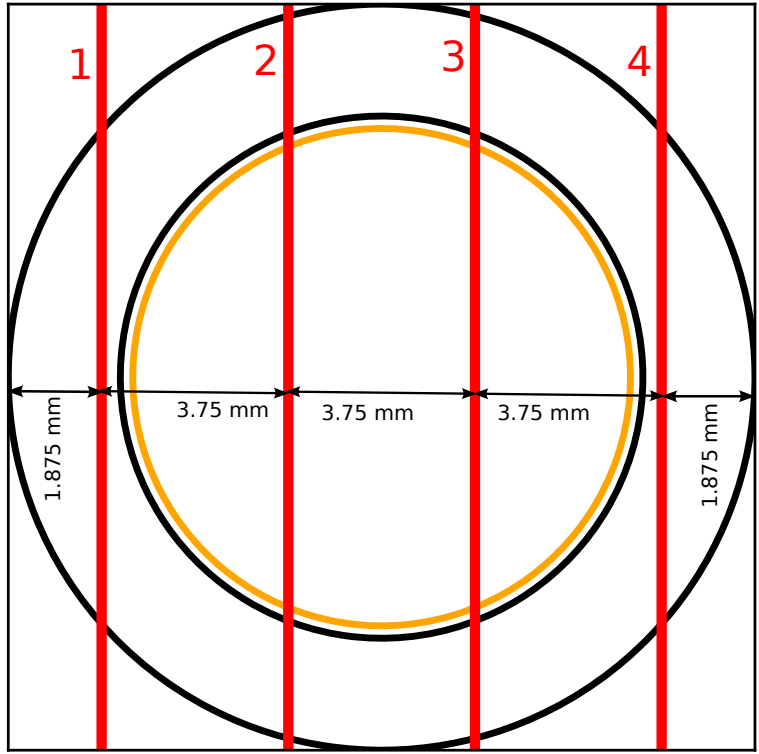}}
	\caption{Multi-slit configuration in VELC. The circle in orange color indicates the disk of the Sun. The inner and outer black colored circles indicate the occulter (1.05 $R_{\odot}$) and FOV of spectroscopy/spectropolarimetry channel (1.5 $R_{\odot}$). The red vertical lines indicate the four slits and are placed symmetrically around the center of the solar image. Folding Mirrors (FM1 and FM2; see Figure \ref{fig:velc}) mounted on a precision Linear Scan Mechanism and that helps in scanning the field on to the slits.}
	\label{fig:slits}
\end{figure*}

In Figure \ref{fig:slits} we show the location of 4 entrance slits of the spectrograph with respect to solar image when linear scan mechanism (LSM) is at Home position (central position). In this position all the 4 slits are symmetrically placed on the image of sun and corona. The image of sun is blocked by the central hole of secondary mirror (M2) of the instrument and makes the image of the corona on the slits only. LSM helps to move the image on the all slits in the horizontal direction and thus any portion of the coronal image can be placed on these slits. The 4 slits forms four spectra around [Fe xiii] 1074.7 nm emission line on the detector. To avoid the overlap of spectra narrow band filter with FWHM of 1.2 nm centered on 1074.7 is kept in front of the detector. The narrow band filter permits to image the spectrum of 3.5 mm width equivalent to 3.8 nm on the detector of 14 mm size. Rest part of the spectrum is blocked by the filter. The separation of 3.75 mm between the slits ensures that there is no mixing of spectra due to one slit with other slit. Further, spectral mask with slot of 1.8 mm width kept at the focus (1074.7 nm) of the spectrograph, to separate out O and E-components of polarized beam will ensure no over lapping of spectra. Thus, the emission spectrum of the whole line profile will be available to determine the polarization and will avoid uncertainty in the determination of magnetic field derived using the part of the emission line recorded using the exit slit \citep{Ulr2009, Loz2015}. In addition, the availability of full emission line profile will also permit to compensate for the Doppler effect. 

It may be noted that instrument polarization will be determined in the laboratory after the integration of the payload before launch. 
In VELC, Fold Mirrors 1 - 4 (FM1, FM2, FM3 and FM4) and diffraction grating operates at large incidence angles to meet the instrument requirements. Large angle of operation can change the polarization states of the reflected beam with respect to the incident beam. In order to estimate the coronal magnetic field accurately, the Stoke’s vectors should be estimated with minimum error from the observed data. To do that, effect of the instrument polarization on the observed data should be estimated. 

Due to the presence of the large angle of operations, state of polarization of the incoming beam (at the entrance aperture) and the recorded data (at the detector) can be different. To estimate this change, system level Muller Matrix will be measured which gives the effect of instrument polarization. A complete details of the calibration set-up and obtained results will be presented as a separate article.

%Complete design of VELC spectropolarimetry channel will be presented elsewhere.

\section{Solar Coronal Magnetic-fields}
\begin{figure*}[!ht]
	\centerline{\includegraphics[width=19cm]{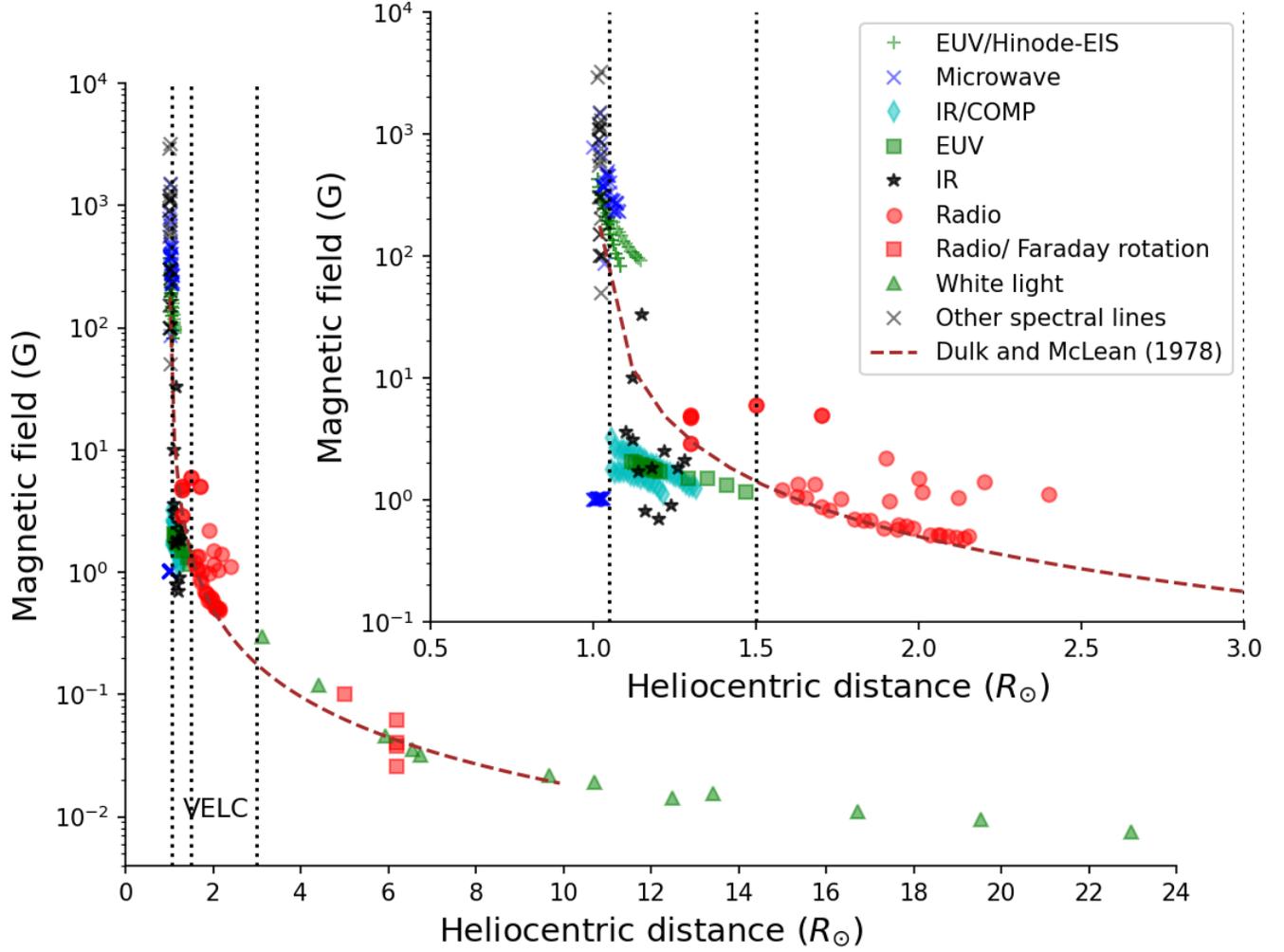}}
	\caption{Summary of solar coronal magnetic fields.The VELC FOV for the continuum channel 1.05 - 3 \Rsun is marked with dotted lines.  The subpanel is the zoomed plot. VELC FOV for both continuum channel (1.05 - 3 \Rsun) and spectropolarimetry channel (1.05 - 1.5 \Rsun) is marked with dotted lines.}
	\label{fig:mf_all}
\end{figure*}

In Figure \ref{fig:mf_all}, we have compiled MF derived by multi wavelength observations in the heliocentric distance from a few Mm (above photosphere) to 24 \Rsun. The vertical dotted lines show the VELC FOV of spectroscopy channel (1.05 - 1.5 \Rsun) and continuum channel (1.05 - 3 \Rsun). The sub-panel shows a zoomed version of magnetic fields in the FOV of VELC. The vertical lines in the sub-panel at 1.05 and 1.5 \Rsun show the FOV of the VELC/SP channel. The vertical lines at 1.05 and 3 \Rsun show the FOV of the VELC continuum channel. It is worth mentioning that previous MF reviews are done by \citet{New1967, New1971, Dulk1978}.

\subsection{Infra-red observations}
In Figure \ref{fig:mf_all}, black markers `*' indicate the MFs derived using the near-infrared coronal emission line Fe XIII 10747 above active regions using spectropolarimeter observations \citep{Lin2000}. The authors have succeeded in observing the weak Stokes V  (circularly polarized intensity) profile from which the line of sight (LOS) MF was measured. Using observations of two active regions, the measured MF strength were 10 and 33 G at heliocentric distance of 1.12 and 1.15 \Rsun, respectively. Similarly \citet{Lin2004} have reported the 0.7 - 3.6 G in the heliocentric distance range 1.10 - 1.28 \Rsun (see black `*' markers in Figure \ref{fig:mf_all}).

\subsection{Hydrogen, Helium and ionized Calcium lines}\label{sec:others}

\citet{Yak2021} presented a simultaneous magnetic field measurements for the limb solar flare using Ca II K, $H_{\delta}$, He I 4471.5 \AA ~and $H_{\beta}$ lines observed on 17 July 1981. Using the Stokes I and Stokes V profiles of the flare observed at two instances of time (that are 16 minutes apart) observed by the Echelle spectrograph of the horizontal solar telescope of the Astronomical Observatory of Taras Shevchenko National University of Kiev, authors derived the coronal MF. In the height range of 10-18 Mm above the photosphere, authors report a strong MF up to 3 kG. Note that these MF are derived using Zeeman splitting method and by using the above mentioned spectral line observations. The MFs derived using this technique are shown as black colored `$\times$' markers in Figure \ref{fig:mf_all}.

\subsection{Polarized microwave emission}
The blue markers show the MFs of three flare loops measured using the polarization equation of a gyrosynchrotron theory \citep{Dulk1985, Sas2014} by \citet{Zhu2020}. Authors used Nobeyama Radioheliograph observations \citep{Nak1994} observed at 17 and 34 GHz. Authors show that MFs varies from $\approx 800$ G near the loop footpoints to $\approx$ 100 G at the height of 10 - 25 Mm, respectively. Using the microwave imaging observations of solar flare of class X8.2 on 2017 September 10 by Expanded
Owens Valley Solar Array (EOVSA), \citet{Chen2020} estimated the MF (see blue colored `$\times$' in Figure \ref{fig:mf_all}) by fitting the microwave spectrum with the gyrosychrotron model. It is worth mentioning that evolution of the MF and magnetic energy with time is discussed by \citet{Fle2020}.

\subsection{Polarization studies of non-thermal radio emission}
Using polarization observations of solar radio bursts (SRBs), it is possible to estimate the MF in solar corona. For example, \citet{Sas2013} used Stokes I (total intensity) and Stokes V (circularly polarized intensity) profiles of harmonic emission of a group of type III bursts and measured the MF. Further, the authors compared those estimates with the one derived using quasi-periodic oscillations of type III bursts. MF measurements derived using both methods are consistent. In another instance, \citep{Sas2014} reported the coronal MF derived using polarization observations of moving type IV burst, which was originated from the core of CME. Authors found that the emission mechanism of the event was optically thin gyrosynchrotron emission and hence used the polarization equation given by \citet{Dulk1985}. These estimates are shown in red markers `o' in Figure \ref{fig:mf_all}. 

\subsection{Band-splitting of solar radio type II bursts}

Magnetohydrodynamic shocks cause solar radio type II bursts in the corona and interplanetary medium (IPM). Most of the time, type II bursts are associated with the CME-driven shocks. Some of the type II bursts often show both fundamental and harmonic emissions. In some cases, both F and H components further show different lanes called band-splitting. \citet{Smerd1974, Smerd1975} shows that such bursts can be used to estimate the magnetic fields. Although many authors reported the MF estimations derived using this method \citep[see, for e.g.,][]{Vrs2002,Cho2007,Vas2014,Har2015,Lat2021}, we used the estimates reported by \citet{Kis2016, Kumari2019}. Such measurements are shown in red colored markers `o' in Figure \ref{fig:mf_all}. 

\subsection{Polarized thermal emission from the `undisturbed' sun}
\citet{Ram2010} reported the magnetic fields in the coronal streamers using low-frequency radio observations. Note that such observations are possible only during rare occasions when the sun is `undisturbed'. Authors estimated the MF strength using the theory outlined by \citet{Sastry2009}. The unpolarized radio emission from the `undisturbed' sun (i.e., streamers) will become polarized in the presence of MF as they split into o-mode and e-mode because of the difference in the optical depths in different directions of the medium. MF estimates using this method are shown in red `o' markers in Figure \ref{fig:mf_all}. 

\subsection{Faraday rotation observations}
The MF in the solar corona and solar wind at the heliocentric distances of $2-15R_{\odot}$ were possible using remote sensing observations. Using a Faraday effect acting on a linearly polarized signal, MFs can be measured while passing through the solar corona \citep{Patzold1987,Spangler2005,Ingleby2007}. For instance, MFs can be estimated by observing a suitable natural radio source or spacecraft beacons during it's occultation \citep{Bird1981,Bird1982}. Such observations were made first time using Pioneer-6 spacecraft \citep{Stelzried1970}. By measuring the orientation vector of the electric field of transmitted and received signals, the Faraday rotation is estimated. In other words, the angle of plane of polarization rotated by the medium is measured from which one can estimate the magnetic field strength of the solar corona.

Based on the observations from the HELIOS spacecraft and using the Faraday rotation technique at a heliocentric distance $3~R_{\odot} \leq R \leq 10 ~R_{\odot}$ MF was estimated by \citet{Patzold1987}. It was about $57$ mG at $6.2~R_{\odot}$. Using the Faraday rotation technique on extragalactic radio sources in the heliocentric distance range $6~ R_{\odot} \leq R \leq 10~ R_{\odot}$, the estimated average magnetic field strength was 40 mG at a distance of $2~ R_{\odot}$ \citep{Spangler2005}. Observations derived using this technique is shown as blue colored `triangles' in Figure \ref{fig:mf_all}. 

\subsection{White light observations: shock standoff distance}\label{sec:shock}

\citet{Gop2011} measured the MF in the heliocentric distance ranging from 6 to 23 \Rsun using the CME observations by white light coronagraphs by measuring the shock standoff distance and radius of curvature of the flux rope. Authors have used the Solar Heliospheric Observatory's (SOHO) C2 and C3 \citep{Bru1995} and COR2 coronagraph of the Sun-Earth
Connection Coronal and Heliospheric Investigation \citep[SEECHI;][]{How2008} on board the Solar Terrestrial Relationship Observatory \citep[STEREO;][]{Kai2008}. The estimated MF vary from 46 mG to 8 mG in the heliocentric distance range 6-23 \Rsun. Assuming adiabatic index, shock Mach number, and \Alfven speeds are measured. Further, polarization brightness observations of SOHO's LASCO/C2 were used to estimate the electron density. Observations derived using this technique are shown as green colored `triangles' in Figure \ref{fig:mf_all}. Similarly \citet{Kum2017} reported the MF using STEREO-A/COR1 and SOHO/LASCO/C2 observations and using shock standoff distance method. Authors reported 0.30 and 0.12 G at the heliocentric distance 3.11 and 4.40 \Rsun, respectively (see green colored triangles in Figure \ref{fig:mf_all}).   

\subsection{EUV observations: shock standoff distance}
\citet{Gop2012}, using CME flux rope and leading shock observations by Solar Dynamic Observatory \citep[SDO;][]{Pes2012} mission's Atmospheric Imaging Assembly \citep[AIA;][]{Lem2012}, \Alfven speed, and then the MF within $\approx 1.4$ \Rsun was measured. In the heliocentric distance range 1.2 - 1.5 \Rsun is varied from 1.3 - 1.5 G. Using the similar observations \citep{Kumari2019} reported the MF to be 1.74 - 1.93 G in the heliocentric distance 1.15 -1.19 \Rsun. These works demonstrate that EUV imaging observations along with the radio type II burst observations can be used as coronal magnetometers \citep{Gop2012}. Observations derived using this technique is shown as green colored `triangles' in Figure \ref{fig:mf_all}. 

\subsection{EUV observations: Magnetic-field induced transitions}
A novel diagnostic technique that allows direct magnetic field measurements from the high spectral resolution observation through so called magnetic-field induced transitions (MIT) effect \citep[][]{Si2020,Landi2020, Lan2021, Bro2021} has recently been used to measure coronal magnetic field. 
Spectroscopic observations of Fe~{\sc x}~and ~{\sc xi} from Hinode/EIS \citep{Cul2007} have been used to estimate the magnetic field in the solar corona by \cite{Landi2020} both on disk and off limb above active regions. The relevant information for the present paper from \cite{Landi2020} is the plot of magnetic field as a function of radial distance. The maximum radial distance for which the field information provided is 1.15 \Rsun. 
%Also, \citet{Bro2021} reported the MF derived using MIT technique using observations of Fe X %257.262 \AA~ from Extreme Ultraviolet Imaging Spectrometer onboard Hinode. In the latter %case, authors report 60 - 150 G. 
The `+' markers shown in Figure \ref{fig:mf_all} indicate the MFs derived using this technique.

\subsection{Observations of waves}
In a recent study \citep{Yang2020a,Yang2020b} has produced a magnetic field map of the solar corona based on the observations of transverse magnetohydrodynamic waves. Using spectroscopic observations with Coronal Multi-Channel Polarimeter they have determined plasma density and the phase speed of the transverse MHD waves. These measurements were combined to produce the coronal magnetic field in the plane of the sky. They have also provided magnetic field as a function of radial distance in a few selected regions of the corona in the height range of 1.05-1.3 \Rsun. In Figure \ref{fig:mf_all}, `diamond' markers in cyan indicate the observations reported by the authors using this technique. 

\subsection{Coronal magnetic field model} 
\citet{Dulk1978} have reviewed the coronal magnetic fields that are reported by then and derived an expression to measure the coronal MF that is valid in the heliocentric distance range 1.02 - 10 \Rsun. The obtained expression is,
 
\begin{equation}\label{eq:B}
B = 0.5 [(R/R_{\odot})-1]^{-1.5}, 
\end{equation}

where R is the heliocentric distance, \Rsun is the solar radii, and B is the MF in G. The above equation is plotted in Figure \ref{fig:mf_all} as a red dotted line. In the sub-panel of Figure \ref{fig:mf_all} it is clear that MFs reported using various techniques are slightly different from the model. 

\section{Sensitivity using Weak Field Approximation}

When the Zeeman splitting ($\Delta \lambda_B$) of a spectral line (Fe XIII) is much
smaller than its typical thermal line broadening ($\Delta \lambda_D$), one can derive a much simpler set of equations that are valid only in the weak magnetic field limit shown below \citep{Cen2018, Fan2018}, 

\begin{equation}\label{wfa}
	V(\lambda) = - K B {dI(\lambda) \over d\lambda}
\end{equation}

where, B is the magnetic field (G), K is the $8.1 \times 10^{-6}~ \rm nm~G^{-1}$ for the Fe XIII line with a Land\'{e} factor of 1.5, and {$dI(\lambda)\over d\lambda$} is first derivative of the intensity profile with respect to wavelength. 

\begin{figure*}[!ht]
	\centerline{\includegraphics[width=15cm]{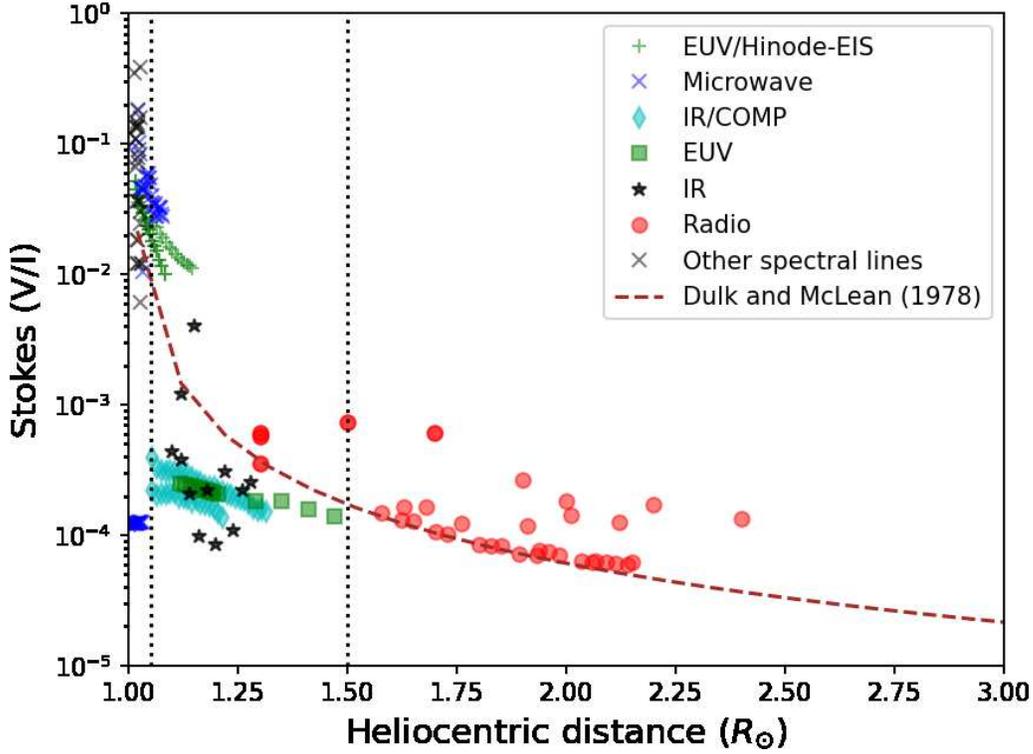}}
	\caption{Measured Stokes V/I under weak field approximation (WFA) in VELC FOV 1.05 - 3 \Rsun. The brown dashed line indicates the measured Stokes V/I assuming the MF model given in equation \ref{eq:B} and WFA. The FOV (1.05 - 1.5 \Rsun) of spectropolarimeter channel is shown using dotted lines.}
	\label{fig:wfa}
\end{figure*}

By knowing the {$dI(\lambda)\over d\lambda$}, we could able to estimate the coronal MF. In this case, we used the {$dI(\lambda)\over d\lambda$} to be 15 that is derived using Norikura observations of the spectral line Fe XIII \citep{Nag2021}, we have derived the required sensitivities (i.e., Stokes V / I intensity) shown in Figure \ref{fig:wfa}. The blue `*', green `+', black `*', cyan `diamond', green `squares', red `circle', and black `$\times$' represents the estimates derived using the observations using microwave, EUV, IR, radio observations, and other spectral line observations (see section \ref{sec:others}) respectively. The brown colored dashed line is derived using the magnetic field model by \citet{Dul1978}. Note that the quantity {$dI(\lambda)\over d\lambda$} may change for different structures present in the corona that needs to be investigated in the future using Fe XIII 1074.7 nm channel observations of VELC.

\section{Summary and Conclusions}
In this article, we have reviewed the recently reported magnetic fields by various direct and indirect techniques. We have also shown that recent reports of MF are not consistent with the MF model by \citet{Dulk1978}. In this article we attempted to constraint the coronal magnetic fields using the reported estimates. For instance in the VELC/SP FOV (1.05 - 1.5 \Rsun) the MFs vary from 0.7 - 203 G. It is worth mentioning that some of the techniques like shock-standoff distance technique discussed in this article (see Section \ref{sec:shock}) can be used to estimate the MF by making use of the VELC continuum channel in the FOV (1.05 - 3 \Rsun).

Using the weak field approximation of Zeeman effect to measure the magnetic field strength we need the Stokes V/I to be in the order ranging from $\approx 6.7 \times 10^{-5}$ to $2 \times 10^{-2}$. Note that with the radially outward direction, coronal field strength decreases and thus one need much lower sensitivities. 

\acknowledgments
KSR acknowledges the VELC team, Dr. K. Nagaraju and Prof. R. Ramesh for their constant support and useful discussions.  We thank all the Scientitists/Engineers at the various
centres of ISRO such as URSC, SAC, LEOS, VSSC etc.
and Indian Institute of Astrophysics who have made great
contributions to the Aditya-L1 mission to reach at the present state.
We gratefully acknowledge the financial support from
ISRO for this project. We thank the referees for providing constructive and valuable suggestions that helped in improving the manuscript.
%% The reference list follows the main body and any appendices.
%% Use LaTeX's thebibliography environment to mark up your reference list.
%% Note \begin{thebibliography} is followed by an empty set of
%% curly braces.  If you forget this, LaTeX will generate the error
%% "Perhaps a missing \item?".
%%
%% thebibliography produces citations in the text using \bibitem-\cite
%% cross-referencing. Each reference is preceded by a
%% \bibitem command that defines in curly braces the KEY that corresponds
%% to the KEY in the \cite commands (see the first section above).
%% Make sure that you provide a unique KEY for every \bibitem or else the
%% paper will not LaTeX. The square brackets should contain
%% the citation text that LaTeX will insert in
%% place of the \cite commands.

%% We have used macros to produce journal name abbreviations.
%% \aastex provides a number of these for the more frequently-cited journals.
%% See the Author Guide for a list of them.

%% Note that the style of the \bibitem labels (in []) is slightly
%% different from previous examples.  The natbib system solves a host
%% of citation expression problems, but it is necessary to clearly
%% delimit the year from the author name used in the citation.
%% See the natbib documentation for more details and options.

\bibliographystyle{aasjournal}
\bibliography{ms}

\begin{thebibliography}{}
\expandafter\ifx\csname natexlab\endcsname\relax\def\natexlab#1{#1}\fi
\providecommand{\url}[1]{\href{#1}{#1}}
\providecommand{\dodoi}[1]{doi:~\href{http://doi.org/#1}{\nolinkurl{#1}}}
\providecommand{\doeprint}[1]{\href{http://ascl.net/#1}{\nolinkurl{http://ascl.net/#1}}}
\providecommand{\doarXiv}[1]{\href{https://arxiv.org/abs/#1}{\nolinkurl{https://arxiv.org/abs/#1}}}

\bibitem[{{Arnaud} \& {Newkirk}(1987)}]{Arnaud1987}
{Arnaud}, J., \& {Newkirk}, Jr., G. 1987, \aap, 178, 263

\bibitem[{{Bird}(1981)}]{Bird1981}
{Bird}, M.~K. 1981, in Solar Wind 4, ed. H.~{Rosenbauer}, 78

\bibitem[{{Bird}(1982)}]{Bird1982}
{Bird}, M.~K. 1982, \ssr, 33, 99, \dodoi{10.1007/BF00213250}

\bibitem[{{B{k{a}}k-St{\c{e}}{\'s}licka}
  {et~al.}(2013){B{k{a}}k-St{\c{e}}{\'s}licka}, {Gibson}, {Fan}, {Bethge},
  {Forland}, \& {Rachmeler}}]{Bak2013}
{B{k{a}}k-St{\c{e}}{\'s}licka}, U., {Gibson}, S.~E., {Fan}, Y., {et~al.} 2013,
  \apjl, 770, L28, \dodoi{10.1088/2041-8205/770/2/L28}

\bibitem[{{Brooks} {et~al.}(2021){Brooks}, {Warren}, \& {Landi}}]{Bro2021}
{Brooks}, D.~H., {Warren}, H.~P., \& {Landi}, E. 2021, \apjl, 915, L24,
  \dodoi{10.3847/2041-8213/ac0c84}

\bibitem[{{Brueckner} {et~al.}(1995){Brueckner}, {Howard}, {Koomen},
  {Korendyke}, {Michels}, {Moses}, {Socker}, {Dere}, {Lamy}, {Llebaria},
  {Bout}, {Schwenn}, {Simnett}, {Bedford}, \& {Eyles}}]{Bru1995}
{Brueckner}, G.~E., {Howard}, R.~A., {Koomen}, M.~J., {et~al.} 1995, \solphys,
  162, 357, \dodoi{10.1007/BF00733434}

\bibitem[{{Carley} {et~al.}(2021){Carley}, {Cecconi}, {Reid}, {Briand},
  {Sasikumar Raja}, {Masson}, {Dorovskyy}, {Tiburzi}, {Vilmer}, {Zucca},
  {Zarka}, {Tagger}, {Griessmeier}, {Corbel}, {Theureau}, {Loh}, \&
  {Girard}}]{Eoin2021}
{Carley}, E.~P., {Cecconi}, B., {Reid}, H.~A., {et~al.} 2021, arXiv e-prints,
  arXiv:2108.05587.
\newblock \doarXiv{2108.05587}

\bibitem[{{Casini} \& {Judge}(1999)}]{Casini1999}
{Casini}, R., \& {Judge}, P.~G. 1999, \apj, 522, 524, \dodoi{10.1086/307629}

\bibitem[{{Centeno}(2018)}]{Cen2018}
{Centeno}, R. 2018, \apj, 866, 89, \dodoi{10.3847/1538-4357/aae087}

\bibitem[{{Chen} {et~al.}(2020){Chen}, {Shen}, {Gary}, {Reeves}, {Fleishman},
  {Yu}, {Guo}, {Krucker}, {Lin}, {Nita}, \& {Kong}}]{Chen2020}
{Chen}, B., {Shen}, C., {Gary}, D.~E., {et~al.} 2020, Nature Astronomy, 4,
  1140, \dodoi{10.1038/s41550-020-1147-7}

\bibitem[{{Chen} {et~al.}(2018){Chen}, {Tian}, {Su}, {Qu}, {Deng}, {Jibben},
  {Yang}, {Zhang}, {Samanta}, {He}, {Wang}, {Zhu}, {Zhong}, \&
  {Liang}}]{Chen2018}
{Chen}, Y., {Tian}, H., {Su}, Y., {et~al.} 2018, \apj, 856, 21,
  \dodoi{10.3847/1538-4357/aaaf68}

\bibitem[{{Cho} {et~al.}(2007){Cho}, {Lee}, {Gary}, {Moon}, \&
  {Park}}]{Cho2007}
{Cho}, K.~S., {Lee}, J., {Gary}, D.~E., {Moon}, Y.~J., \& {Park}, Y.~D. 2007,
  \apj, 665, 799, \dodoi{10.1086/519160}

\bibitem[{{Culhane} {et~al.}(2007){Culhane}, {Harra}, {James}, {Al-Janabi},
  {Bradley}, {Chaudry}, {Rees}, {Tandy}, {Thomas}, {Whillock}, {Winter},
  {Doschek}, {Korendyke}, {Brown}, {Myers}, {Mariska}, {Seely}, {Lang}, {Kent},
  {Shaughnessy}, {Young}, {Simnett}, {Castelli}, {Mahmoud}, {Mapson-Menard},
  {Probyn}, {Thomas}, {Davila}, {Dere}, {Windt}, {Shea}, {Hagood}, {Moye},
  {Hara}, {Watanabe}, {Matsuzaki}, {Kosugi}, {Hansteen}, \&
  {Wikstol}}]{Cul2007}
{Culhane}, J.~L., {Harra}, L.~K., {James}, A.~M., {et~al.} 2007, \solphys, 243,
  19, \dodoi{10.1007/s01007-007-0293-1}

\bibitem[{{Dulk}(1985)}]{Dulk1985}
{Dulk}, G.~A. 1985, \araa, 23, 169, \dodoi{10.1146/annurev.aa.23.090185.001125}

\bibitem[{{Dulk} \& {McLean}(1978{\natexlab{a}})}]{Dulk1978}
{Dulk}, G.~A., \& {McLean}, D.~J. 1978{\natexlab{a}}, \solphys, 57, 279,
  \dodoi{10.1007/BF00160102}

\bibitem[{{Dulk} \& {McLean}(1978{\natexlab{b}})}]{Dul1978}
---. 1978{\natexlab{b}}, \solphys, 57, 279, \dodoi{10.1007/BF00160102}

\bibitem[{{Fan} {et~al.}(2018){Fan}, {Gibson}, \& {Tomczyk}}]{Fan2018}
{Fan}, Y., {Gibson}, S., \& {Tomczyk}, S. 2018, \apj, 866, 57,
  \dodoi{10.3847/1538-4357/aadd0e}

\bibitem[{{Fleishman} {et~al.}(2020){Fleishman}, {Gary}, {Chen}, {Kuroda},
  {Yu}, \& {Nita}}]{Fle2020}
{Fleishman}, G.~D., {Gary}, D.~E., {Chen}, B., {et~al.} 2020, Science, 367,
  278, \dodoi{10.1126/science.aax6874}

\bibitem[{{Gibson} {et~al.}(2016){Gibson}, {Kucera}, {White}, {Dove}, {Fan},
  {Forland}, {Rachmeler}, {Downs}, \& {Reeves}}]{Gib2016}
{Gibson}, S., {Kucera}, T., {White}, S., {et~al.} 2016, Frontiers in Astronomy
  and Space Sciences, 3, 8, \dodoi{10.3389/fspas.2016.00008}

\bibitem[{{Gibson} {et~al.}(2017){Gibson}, {Dalmasse}, {Rachmeler}, {De Rosa},
  {Tomczyk}, {de Toma}, {Burkepile}, \& {Galloy}}]{Gib2017}
{Gibson}, S.~E., {Dalmasse}, K., {Rachmeler}, L.~A., {et~al.} 2017, \apjl, 840,
  L13, \dodoi{10.3847/2041-8213/aa6fac}

\bibitem[{{Gopalswamy} {et~al.}(2012){Gopalswamy}, {Nitta}, {Akiyama},
  {M{\"a}kel{\"a}}, \& {Yashiro}}]{Gop2012}
{Gopalswamy}, N., {Nitta}, N., {Akiyama}, S., {M{\"a}kel{\"a}}, P., \&
  {Yashiro}, S. 2012, \apj, 744, 72, \dodoi{10.1088/0004-637X/744/1/72}

\bibitem[{{Gopalswamy} \& {Yashiro}(2011)}]{Gop2011}
{Gopalswamy}, N., \& {Yashiro}, S. 2011, \apjl, 736, L17,
  \dodoi{10.1088/2041-8205/736/1/L17}

\bibitem[{{Hale}(1908)}]{Hale1908}
{Hale}, G.~E. 1908, \apj, 28, 315, \dodoi{10.1086/141602}

\bibitem[{{Hariharan} {et~al.}(2015){Hariharan}, {Ramesh}, \&
  {Kathiravan}}]{Har2015}
{Hariharan}, K., {Ramesh}, R., \& {Kathiravan}, C. 2015, \solphys, 290, 2479,
  \dodoi{10.1007/s11207-015-0761-5}

\bibitem[{{Hariharan} {et~al.}(2014){Hariharan}, {Ramesh}, {Kishore},
  {Kathiravan}, \& {Gopalswamy}}]{Har2014}
{Hariharan}, K., {Ramesh}, R., {Kishore}, P., {Kathiravan}, C., \&
  {Gopalswamy}, N. 2014, \apj, 795, 14, \dodoi{10.1088/0004-637X/795/1/14}

\bibitem[{{Harvey}(1969)}]{Harvey1969}
{Harvey}, J.~W. 1969, PhD thesis, Colorado Univ., Boulder.

\bibitem[{{Howard} {et~al.}(2008){Howard}, {Moses}, {Vourlidas}, {Newmark},
  {Socker}, {Plunkett}, {Korendyke}, {Cook}, {Hurley}, {Davila}, {Thompson},
  {St Cyr}, {Mentzell}, {Mehalick}, {Lemen}, {Wuelser}, {Duncan}, {Tarbell},
  {Wolfson}, {Moore}, {Harrison}, {Waltham}, {Lang}, {Davis}, {Eyles},
  {Mapson-Menard}, {Simnett}, {Halain}, {Defise}, {Mazy}, {Rochus}, {Mercier},
  {Ravet}, {Delmotte}, {Auchere}, {Delaboudiniere}, {Bothmer}, {Deutsch},
  {Wang}, {Rich}, {Cooper}, {Stephens}, {Maahs}, {Baugh}, {McMullin}, \&
  {Carter}}]{How2008}
{Howard}, R.~A., {Moses}, J.~D., {Vourlidas}, A., {et~al.} 2008, \ssr, 136, 67,
  \dodoi{10.1007/s11214-008-9341-4}

\bibitem[{{Ingleby} {et~al.}(2007){Ingleby}, {Spangler}, \&
  {Whiting}}]{Ingleby2007}
{Ingleby}, L.~D., {Spangler}, S.~R., \& {Whiting}, C.~A. 2007, \apj, 668, 520,
  \dodoi{10.1086/521140}

\bibitem[{{Jibben} {et~al.}(2016){Jibben}, {Reeves}, \& {Su}}]{Jib2016}
{Jibben}, P., {Reeves}, K., \& {Su}, Y. 2016, Frontiers in Astronomy and Space
  Sciences, 3, 10, \dodoi{10.3389/fspas.2016.00010}

\bibitem[{{Kaiser} {et~al.}(2008){Kaiser}, {Kucera}, {Davila}, {St. Cyr},
  {Guhathakurta}, \& {Christian}}]{Kai2008}
{Kaiser}, M.~L., {Kucera}, T.~A., {Davila}, J.~M., {et~al.} 2008, \ssr, 136, 5,
  \dodoi{10.1007/s11214-007-9277-0}

\bibitem[{{Kishore} {et~al.}(2017){Kishore}, {Kathiravan}, {Ramesh}, \&
  {Ebenezer}}]{Kis2017}
{Kishore}, P., {Kathiravan}, C., {Ramesh}, R., \& {Ebenezer}, E. 2017, Journal
  of Astrophysics and Astronomy, 38, 24, \dodoi{10.1007/s12036-017-9444-y}

\bibitem[{{Kishore} {et~al.}(2016){Kishore}, {Ramesh}, {Hariharan},
  {Kathiravan}, \& {Gopalswamy}}]{Kis2016}
{Kishore}, P., {Ramesh}, R., {Hariharan}, K., {Kathiravan}, C., \&
  {Gopalswamy}, N. 2016, \apj, 832, 59, \dodoi{10.3847/0004-637X/832/1/59}

\bibitem[{{Kuhn}(1995)}]{Kuhn1995}
{Kuhn}, J.~R. 1995, in Infrared tools for solar astrophysics: What's next?, ed.
  J.~R. {Kuhn} \& M.~J. {Penn}, 89

\bibitem[{{Kumari} {et~al.}(2017{\natexlab{a}}){Kumari}, {Ramesh},
  {Kathiravan}, \& {Wang}}]{Kumari2017}
{Kumari}, A., {Ramesh}, R., {Kathiravan}, C., \& {Wang}, T.~J.
  2017{\natexlab{a}}, \solphys, 292, 161, \dodoi{10.1007/s11207-017-1180-6}

\bibitem[{{Kumari} {et~al.}(2017{\natexlab{b}}){Kumari}, {Ramesh},
  {Kathiravan}, \& {Wang}}]{Kum2017}
---. 2017{\natexlab{b}}, \solphys, 292, 161, \dodoi{10.1007/s11207-017-1180-6}

\bibitem[{{Kumari} {et~al.}(2019){Kumari}, {Ramesh}, {Kathiravan}, {Wang}, \&
  {Gopalswamy}}]{Kumari2019}
{Kumari}, A., {Ramesh}, R., {Kathiravan}, C., {Wang}, T.~J., \& {Gopalswamy},
  N. 2019, \apj, 881, 24, \dodoi{10.3847/1538-4357/ab2adf}

\bibitem[{{Landi} {et~al.}(2020){Landi}, {Hutton}, {Brage}, \&
  {Li}}]{Landi2020}
{Landi}, E., {Hutton}, R., {Brage}, T., \& {Li}, W. 2020, \apj, 904, 87,
  \dodoi{10.3847/1538-4357/abbf54}

\bibitem[{{Landi} {et~al.}(2021){Landi}, {Li}, {Brage}, \& {Hutton}}]{Lan2021}
{Landi}, E., {Li}, W., {Brage}, T., \& {Hutton}, R. 2021, \apj, 913, 1,
  \dodoi{10.3847/1538-4357/abf6d1}

\bibitem[{{Lata Soni} {et~al.}(2021){Lata Soni}, {Ebenezer}, \& {lal
  Yadav}}]{Lat2021}
{Lata Soni}, S., {Ebenezer}, E., \& {lal Yadav}, M. 2021, \apss, 366, 31,
  \dodoi{10.1007/s10509-021-03933-7}

\bibitem[{{Lemen} {et~al.}(2012){Lemen}, {Title}, {Akin}, {Boerner}, {Chou},
  {Drake}, {Duncan}, {Edwards}, {Friedlaender}, {Heyman}, {Hurlburt}, {Katz},
  {Kushner}, {Levay}, {Lindgren}, {Mathur}, {McFeaters}, {Mitchell}, {Rehse},
  {Schrijver}, {Springer}, {Stern}, {Tarbell}, {Wuelser}, {Wolfson}, {Yanari},
  {Bookbinder}, {Cheimets}, {Caldwell}, {Deluca}, {Gates}, {Golub}, {Park},
  {Podgorski}, {Bush}, {Scherrer}, {Gummin}, {Smith}, {Auker}, {Jerram},
  {Pool}, {Soufli}, {Windt}, {Beardsley}, {Clapp}, {Lang}, \&
  {Waltham}}]{Lem2012}
{Lemen}, J.~R., {Title}, A.~M., {Akin}, D.~J., {et~al.} 2012, \solphys, 275,
  17, \dodoi{10.1007/s11207-011-9776-8}

\bibitem[{{Lin} {et~al.}(2004){Lin}, {Kuhn}, \& {Coulter}}]{Lin2004}
{Lin}, H., {Kuhn}, J.~R., \& {Coulter}, R. 2004, \apjl, 613, L177,
  \dodoi{10.1086/425217}

\bibitem[{{Lin} {et~al.}(2000){Lin}, {Penn}, \& {Tomczyk}}]{Lin2000}
{Lin}, H., {Penn}, M.~J., \& {Tomczyk}, S. 2000, \apjl, 541, L83,
  \dodoi{10.1086/312900}

\bibitem[{{Lozitsky}(2015)}]{Loz2015}
{Lozitsky}, V.~G. 2015, Advances in Space Research, 55, 958,
  \dodoi{10.1016/j.asr.2014.09.028}

\bibitem[{{Mickey}(1973)}]{Mickey1973}
{Mickey}, D.~L. 1973, \apjl, 181, L19, \dodoi{10.1086/181175}

\bibitem[{{Nagaraju} {et~al.}(2021){Nagaraju}, {Prasad}, {Hegde}, {Narra},
  {Utkarsha}, {Kumar}, {Singh}, \& {Kumar}}]{Nag2021}
{Nagaraju}, K., {Prasad}, B.~R., {Hegde}, B.~S., {et~al.} 2021, \ao, 60, 8145,
  \dodoi{10.1364/AO.434219}

\bibitem[{{Nakajima} {et~al.}(1994){Nakajima}, {Nishio}, {Enome}, {Shibasaki},
  {Takano}, {Hanaoka}, {Torii}, {Sekiguchi}, {Bushimata}, {Kawashima},
  {Shinohara}, {Irimajiri}, {Koshiishi}, {Kosugi}, {Shiomi}, {Sawa}, \&
  {Kai}}]{Nak1994}
{Nakajima}, H., {Nishio}, M., {Enome}, S., {et~al.} 1994, IEEE Proceedings, 82,
  705

\bibitem[{{Narra} {et~al.}(2020){Narra}, {Budihal}, {Somasundaram}, {Sriraman},
  {Venkatasubramanian}, {Padavu}, {Mani}, {Basavaraja}, \& {Naik}}]{Narra2020}
{Narra}, V.~S., {Budihal}, R.~P., {Somasundaram}, P.~K., {et~al.} 2020,
  Experimental Astronomy, 50, 265, \dodoi{10.1007/s10686-020-09675-8}

\bibitem[{{Ndacyayisenga} {et~al.}(2021){Ndacyayisenga}, {Uwamahoro},
  {Sasikumar Raja}, \& {Monstein}}]{Nda2021}
{Ndacyayisenga}, T., {Uwamahoro}, J., {Sasikumar Raja}, K., \& {Monstein}, C.
  2021, Advances in Space Research, 67, 1425, \dodoi{10.1016/j.asr.2020.11.022}

\bibitem[{{Newkirk}(1971)}]{New1971}
{Newkirk}, G., J. 1971, {Coronal Magnetic Fields}, ed. C.~J. {Macris}, Vol.~27,
  66, \dodoi{10.1007/978-90-277-0204-3\_5}

\bibitem[{{Newkirk}(1967)}]{New1967}
{Newkirk}, Gordon, J. 1967, \araa, 5, 213,
  \dodoi{10.1146/annurev.aa.05.090167.001241}

\bibitem[{{Patzold} {et~al.}(1987){Patzold}, {Bird}, {Volland}, {Levy},
  {Seidel}, \& {Stelzried}}]{Patzold1987}
{Patzold}, M., {Bird}, M.~K., {Volland}, H., {et~al.} 1987, \solphys, 109, 91,
  \dodoi{10.1007/BF00167401}

\bibitem[{{Pesnell} {et~al.}(2012){Pesnell}, {Thompson}, \&
  {Chamberlin}}]{Pes2012}
{Pesnell}, W.~D., {Thompson}, B.~J., \& {Chamberlin}, P.~C. 2012, \solphys,
  275, 3, \dodoi{10.1007/s11207-011-9841-3}

\bibitem[{{Querfeld} \& {Smartt}(1984)}]{Querfeld1984}
{Querfeld}, C.~W., \& {Smartt}, R.~N. 1984, \solphys, 91, 299,
  \dodoi{10.1007/BF00146301}

\bibitem[{{Raghavendra Prasad} {et~al.}(2017){Raghavendra Prasad}, {Banerjee},
  {Singh}, {Nagabhushana}, {Kumar}, {Kamath}, {Kathiravan}, {Venkata},
  {Rajkumar}, {Natarajan}, {Juneja}, {Somu}, {Pant}, {Shaji},
  {Sankarsubramanian}, {Patra}, {Venkateswaran}, {Adoni}, {Narendra},
  {Haridas}, {Mathew}, {Mohan Krishna}, {Amareswari}, \&
  {Jaiswal}}]{Prasad2017}
{Raghavendra Prasad}, B., {Banerjee}, D., {Singh}, J., {et~al.} 2017, Current
  Science, 113, 613

\bibitem[{{Raj Kumar} {et~al.}(2018){Raj Kumar}, {Raghavendra Prasad}, {Singh},
  \& {Venkata}}]{Raj2018}
{Raj Kumar}, N., {Raghavendra Prasad}, B., {Singh}, J., \& {Venkata}, S. 2018,
  Experimental Astronomy, 45, 219, \dodoi{10.1007/s10686-017-9569-7}

\bibitem[{{Ramesh} {et~al.}(2010){Ramesh}, {Kathiravan}, \& {Sastry}}]{Ram2010}
{Ramesh}, R., {Kathiravan}, C., \& {Sastry}, C.~V. 2010, \apj, 711, 1029,
  \dodoi{10.1088/0004-637X/711/2/1029}

\bibitem[{{Sasikumar Raja} {et~al.}(2019){Sasikumar Raja}, {Janardhan},
  {Bisoi}, {Ingale}, {Subramanian}, {Fujiki}, \& {Maksimovic}}]{Sas2019}
{Sasikumar Raja}, K., {Janardhan}, P., {Bisoi}, S.~K., {et~al.} 2019, \solphys,
  294, 123, \dodoi{10.1007/s11207-019-1514-7}

\bibitem[{{Sasikumar Raja} \& {Ramesh}(2013)}]{Sas2013}
{Sasikumar Raja}, K., \& {Ramesh}, R. 2013, \apj, 775, 38,
  \dodoi{10.1088/0004-637X/775/1/38}

\bibitem[{{Sasikumar Raja} {et~al.}(2014){Sasikumar Raja}, {Ramesh},
  {Hariharan}, {Kathiravan}, \& {Wang}}]{Sas2014}
{Sasikumar Raja}, K., {Ramesh}, R., {Hariharan}, K., {Kathiravan}, C., \&
  {Wang}, T.~J. 2014, \apj, 796, 56, \dodoi{10.1088/0004-637X/796/1/56}

\bibitem[{{Sastry}(2009)}]{Sastry2009}
{Sastry}, C.~V. 2009, \apj, 697, 1934, \dodoi{10.1088/0004-637X/697/2/1934}

\bibitem[{{Schatten} {et~al.}(1969){Schatten}, {Wilcox}, \&
  {Ness}}]{Schatten1969}
{Schatten}, K.~H., {Wilcox}, J.~M., \& {Ness}, N.~F. 1969, \solphys, 6, 442,
  \dodoi{10.1007/BF00146478}

\bibitem[{{Schrijver} \& {De Rosa}(2003)}]{Sch2003}
{Schrijver}, C.~J., \& {De Rosa}, M.~L. 2003, \solphys, 212, 165,
  \dodoi{10.1023/A:1022908504100}

\bibitem[{{Seetha} \& {Megala}(2017)}]{Seetha2017}
{Seetha}, S., \& {Megala}, S. 2017, Current Science, 113, 610

\bibitem[{{Si} {et~al.}(2020){Si}, {Brage}, {Li}, {Grumer}, {Li}, \&
  {Hutton}}]{Si2020}
{Si}, R., {Brage}, T., {Li}, W., {et~al.} 2020, \apjl, 898, L34,
  \dodoi{10.3847/2041-8213/aba18c}

\bibitem[{{Singh} {et~al.}(2019{\natexlab{a}}){Singh}, {Sasikumar Raja},
  {Subramanian}, {Ramesh}, \& {Monstein}}]{Dayal2019}
{Singh}, D., {Sasikumar Raja}, K., {Subramanian}, P., {Ramesh}, R., \&
  {Monstein}, C. 2019{\natexlab{a}}, \solphys, 294, 112,
  \dodoi{10.1007/s11207-019-1500-0}

\bibitem[{{Singh} {et~al.}(2019{\natexlab{b}}){Singh}, {Prasad}, {Venkata}, \&
  {Kumar}}]{Singh2019}
{Singh}, J., {Prasad}, B.~R., {Venkata}, S., \& {Kumar}, A. 2019{\natexlab{b}},
  Advances in Space Research, 64, 1455, \dodoi{10.1016/j.asr.2019.07.007}

\bibitem[{{Smerd} {et~al.}(1974){Smerd}, {Sheridan}, \& {Stewart}}]{Smerd1974}
{Smerd}, S.~F., {Sheridan}, K.~V., \& {Stewart}, R.~T. 1974, in Coronal
  Disturbances, ed. G.~A. {Newkirk}, Vol.~57, 389

\bibitem[{{Smerd} {et~al.}(1975){Smerd}, {Sheridan}, \& {Stewart}}]{Smerd1975}
{Smerd}, S.~F., {Sheridan}, K.~V., \& {Stewart}, R.~T. 1975, \aplett, 16, 23

\bibitem[{{Spangler}(2005)}]{Spangler2005}
{Spangler}, S.~R. 2005, \ssr, 121, 189, \dodoi{10.1007/s11214-006-4719-7}

\bibitem[{{Stelzried} {et~al.}(1970){Stelzried}, {Levy}, {Sato}, {Rusch},
  {Ohlson}, {Schatten}, \& {Wilcox}}]{Stelzried1970}
{Stelzried}, C.~T., {Levy}, G.~S., {Sato}, T., {et~al.} 1970, \solphys, 14,
  440, \dodoi{10.1007/BF00221330}

\bibitem[{{Ulrich} {et~al.}(2009){Ulrich}, {Bertello}, {Boyden}, \&
  {Webster}}]{Ulr2009}
{Ulrich}, R.~K., {Bertello}, L., {Boyden}, J.~E., \& {Webster}, L. 2009,
  \solphys, 255, 53, \dodoi{10.1007/s11207-008-9302-9}

\bibitem[{{Umuhire} {et~al.}(2021){Umuhire}, {Uwamahoro}, {Sasikumar Raja},
  {Kumari}, \& {Monstein}}]{Umi2021}
{Umuhire}, A.~C., {Uwamahoro}, J., {Sasikumar Raja}, K., {Kumari}, A., \&
  {Monstein}, C. 2021, Advances in Space Research, 68, 3464,
  \dodoi{10.1016/j.asr.2021.06.029}

\bibitem[{{Vasanth} {et~al.}(2014){Vasanth}, {Umapathy}, {Vr{\v{s}}nak},
  {{\v{Z}}ic}, \& {Prakash}}]{Vas2014}
{Vasanth}, V., {Umapathy}, S., {Vr{\v{s}}nak}, B., {{\v{Z}}ic}, T., \&
  {Prakash}, O. 2014, \solphys, 289, 251, \dodoi{10.1007/s11207-013-0318-4}

\bibitem[{{Vemareddy} {et~al.}(2013){Vemareddy}, {Ambastha}, \&
  {Wiegelmann}}]{Vema2013}
{Vemareddy}, P., {Ambastha}, A., \& {Wiegelmann}, T. 2013, Bulletin of the
  Astronomical Society of India, 41, 183.
\newblock \doarXiv{1310.6895}

\bibitem[{Venkata {et~al.}(2020)Venkata, Budihal, Venkatasubramanian, Padavu,
  \& Sriraman}]{SV2020}
Venkata, S., Budihal, R.~P., Venkatasubramanian, N., Padavu, U.~K., \&
  Sriraman, K. 2020, Optical Engineering, 59, 1 ,
  \dodoi{10.1117/1.OE.59.8.084106}

\bibitem[{Venkata {et~al.}(2017)Venkata, Prasad, Nalla, \& Singh}]{SV2017}
Venkata, S.~N., Prasad, B.~R., Nalla, R.~K., \& Singh, J. 2017, Journal of
  Astronomical Telescopes, Instruments, and Systems, 3,
  \dodoi{10.1117/1.JATIS.3.1.014002}

\bibitem[{{Vr{\v{s}}nak} {et~al.}(2002){Vr{\v{s}}nak}, {Magdaleni{\'c}},
  {Aurass}, \& {Mann}}]{Vrs2002}
{Vr{\v{s}}nak}, B., {Magdaleni{\'c}}, J., {Aurass}, H., \& {Mann}, G. 2002,
  \aap, 396, 673, \dodoi{10.1051/0004-6361:20021413}

\bibitem[{{Wiegelmann} \& {Sakurai}(2012)}]{Wie2012}
{Wiegelmann}, T., \& {Sakurai}, T. 2012, Living Reviews in Solar Physics, 9, 5,
  \dodoi{10.12942/lrsp-2012-5}

\bibitem[{{Wiegelmann} {et~al.}(2014){Wiegelmann}, {Thalmann}, \&
  {Solanki}}]{Wie2014}
{Wiegelmann}, T., {Thalmann}, J.~K., \& {Solanki}, S.~K. 2014, \aapr, 22, 78,
  \dodoi{10.1007/s00159-014-0078-7}

\bibitem[{{Yakovkin} {et~al.}(2021){Yakovkin}, {Veronig}, \&
  {Lozitsky}}]{Yak2021}
{Yakovkin}, I.~I., {Veronig}, A.~M., \& {Lozitsky}, V.~G. 2021, Advances in
  Space Research, 68, 1507, \dodoi{10.1016/j.asr.2021.03.036}

\bibitem[{{Yang} {et~al.}({2020b}){Yang}, {Tian}, {Tomczyk}, {Morton}, {Bai},
  {Samanta}, \& {Chen}}]{Yang2020b}
{Yang}, Z., {Tian}, H., {Tomczyk}, S., {et~al.} {2020b}, Science in China E:
  Technological Sciences, 63, 2357, \dodoi{10.1007/s11431-020-1706-9}

\bibitem[{{Yang} {et~al.}({2020a}){Yang}, {Bethge}, {Tian}, {Tomczyk},
  {Morton}, {Del Zanna}, {McIntosh}, {Karak}, {Gibson}, {Samanta}, {He},
  {Chen}, \& {Wang}}]{Yang2020a}
{Yang}, Z., {Bethge}, C., {Tian}, H., {et~al.} {2020a}, Science, 369, 694,
  \dodoi{10.1126/science.abb4462}

\bibitem[{{Zhu} {et~al.}(2020){Zhu}, {Tan}, {Su}, {Tian}, {Xu}, {Chen}, {Song},
  \& {Tan}}]{Zhu2020}
{Zhu}, R., {Tan}, B., {Su}, Y., {et~al.} 2020, arXiv e-prints,
  arXiv:2006.15014.
\newblock \doarXiv{2006.15014}

\end{thebibliography}

%% This command is needed to show the entire author+affilation list when
%% the collaboration and author truncation commands are used.  It has to
%% go at the end of the manuscript.
%\allauthors

%% Include this line if you are using the \added, \replaced, \deleted
%% commands to see a summary list of all changes at the end of the article.
%\listofchanges

\end{document}